\documentclass[12pt]{iopart}
\usepackage{epsf}
\usepackage{epsfig}
\usepackage{amssymb}

\begin{document}

\title{Ground-state energy and entropy of the two-dimensional Edwards-Anderson spin-glass model with different bond distributions}

\author{D. J. Perez-Morelo$^1$, A. J. Ramirez-Pastor$^2$ and F. Rom\'a$^2$}
\address{$^1$ Centro At{\'{o}}mico Bariloche, CONICET, San Carlos de
Bariloche, R8402AGP R\'{\i}o Negro, Argentina. \\
$^2$ Departamento de F\'{\i}sica, INFAP, Universidad Nacional de
San Luis, CONICET, Chacabuco 917, D5700BWS San Luis, Argentina.}

\date{\today}

\begin{abstract}
We study the two-dimensional Edwards-Anderson spin-glass model using a parallel tempering Monte Carlo algorithm.  The ground-state energy and entropy are calculated for different bond distributions.  In particular, the entropy is obtained by using a thermodynamic integration technique and an appropriate reference state, which is determined with the method of high-temperature expansion. This strategy provide accurate values of this quantity for finite-size lattices. By extrapolating to the thermodynamic limit, the ground-state energy and entropy of the different versions of the spin-glass model are determined.

\end{abstract}

\pacs{Valid PACS appear here}

\maketitle

\section{\label{Intro}Introduction}

The study of disordered and frustrated systems is a current subject in statistical mechanics, and spin-glass models occupy a privileged place \cite{Bovier,Mezard,FischerHertz,Young}.  At low temperatures and even in the ground state (GS), these complex systems display the main characteristics which dominate its physical behavior.  GS quantities such as the domain-walls energy, for example, are frequently calculated to determine if a finite critical temperature exists \cite{Bray1984,McMillan1984,Hartmann1999,Hartmann2001a,Roma2007,Risau2008}.  

The energy and entropy are other GS observables that give valuable information of these systems.  The former can be calculated using different optimization techniques \cite{Hartmannlibro1,Hartmannlibro2}, among others, genetic algorithms \cite{Holland1975}, simulated annealing \cite{Kirkpatrick1983}, multicanonical ensemble \cite{Berg1994} and parallel tempering \cite{Geyer1991,Hukushima1996,Moreno2003,Roma2009}. On the other hand, as entropy calculations are more difficult to carry out, sophisticated algorithms have been designed to determine this quantity.  For example, transfer matrix \cite{Morgenstern1980,Cheung1983,Kolan1982} and ballistic-search \cite{Hartmann2001b} methods.  A more popular technique to calculate entropy is known as the {\em thermodynamic integration method} \cite{Kirkpatrick1977,Binder1985,Roma2004}. This relies upon integration of the internal energy as function of temperature along a reversible path. Initial point corresponds to an arbitrary but known {\em reference state}, while the final point corresponds to the state for which the entropy value is required.  In the practice, a problem to implement this technique is the necessity to calculate a suitable reference entropy at a not very high temperature.  

In this work we use a parallel tempering algorithm to determine both the GS energy and, by means of the thermodynamic integration method, the GS entropy of the two-dimensional (2D) Edwards-Anderson spin-glass model \cite{EA}, a paradigmatic disordered and frustrated system.   By considering different bond distributions, the more efficient strategies to implement the thermodynamic integration method are determined in each case.  These strategies are based on the construction of reference states by the {\em method of high-temperature expansion}.   We stress that, the main objective of this work, is to show that this implementation of the thermodynamic integration technique permits to obtain reliable values of the GS entropy for different disordered and frustrated models.  

The article is organized as follows.  In Section \ref{ModAlg} the Edwards-Anderson spin-glass model and the parallel tempering algorithm are presented.  The thermodynamic integration method is described in Section \ref{TIM} and also the construction of reference states is discussed. In Section \ref{NumRes}, we present our main results and the values of GS energy and entropy obtained by extrapolating to the thermodynamic limit from finite-lattice calculations.  Conclusions are summarized in Section \ref{Conc}.   

\section{\label{ModAlg}Model and algorithm}

We start by considering the Hamiltonian of the $d$-dimensional Edwards-Anderson spin-glass model \cite{EA},
\begin{equation}
H = - \sum_{(ij)} J_{ij} \sigma_i \sigma_j, \label{ham}
\end{equation}
where the sum runs over the nearest neighbors of a hypercubic lattice of linear dimension $L$ and the variables $\sigma_i = \pm 1$ represent to $N=L^d$ Ising spins.  The coupling constants or bonds, $J_{ij}$, are independent random variables chosen from a given probability distribution.  Along this work, all 2D samples (particular realizations of bond disorder) with a square geometry were generated with periodic-periodic boundary conditions.

We study the Edwards-Anderson model with different bond distributions.  On one hand, we consider the typical continuous Gaussian,
\begin{equation}
P_{\mathrm{G}}(J_{ij})= \frac{1}{\sqrt{2 \pi}} \exp(- J_{ij}^2/2), \label{DistG}
\end{equation}
and discrete bimodal,
\begin{equation}
P_{\mathrm{B}}(J_{ij})=\frac{1}{2} \left[\delta(J_{ij}-1)+\delta(J_{ij}+1)\right],   \label{DistB}
\end{equation}
distributions, for which the mean value is zero and the variance is one.  These are the most popular bond distributions.  To avoid confusions, we will denominate EAG and EAB to the versions of the Edwards-Anderson model where interactions are drawn from, respectively, equations (\ref{DistG}) and (\ref{DistB}).  

On the other hand, we also study a continuous uniform distribution with zero-mean value and variance one,      
\begin{equation}
P_{\mathrm{U}}(J_{ij})= \left\{ 
\begin{array}{ll}
1/(2 \sqrt{3})  & \ \ \textrm{if $|J_{ij}| \le \sqrt{3}$}\\
0               & \ \ \textrm{if $|J_{ij}|  >  \sqrt{3}$},   \label{DistU}
\end{array} 
\right.
\end{equation}
and a $2p$-delta distribution,
\begin{equation}
P_{\mathrm{P}}(J_{ij})=\frac{1}{2p} \sum_{i=1}^p \left[\delta(J_{ij}-ib)+\delta(J_{ij}+ib)\right],   \label{DistP}
\end{equation}
which is a generalization of (\ref{DistB}).  Here, $b(p)$ is a function of $p$ chosen so that the mean value and the variance of $J_{ij}$ to be zero and one, respectively.  By direct integration it is easy to show that
\begin{equation} 
b(p)=\sqrt{\frac{6}{(p+1)(2p+1)}}.
\end{equation} 
For $p=1$ both distributions (\ref{DistB}) and (\ref{DistP}) are equivalent.  On the other hand, $b_{\textrm{max}} \equiv p b\to \sqrt{3}$ (the maximum possible value of $J_{ij}$) when $p \to \infty$ and the equation (\ref{DistP}) tend to the uniform distribution (\ref{DistU}).  

In addition, an asymmetric distribution \cite{Shirakura1997},
\begin{equation}
P_{\mathrm{A}}(J_{ij})=\frac{1}{2} \left[\delta(J_{ij}-1)+\delta(J_{ij}+a)\right],   \label{DistA}
\end{equation}
where $0 \le a \le 1$ is a rational number, and an irrational distribution \cite{Jorg2006},
\begin{equation}
P_{\mathrm{I}}(J_{ij})=\frac{1}{4} \left[\delta(J_{ij} \pm 1)+\delta(J_{ij} \pm c)\right],   \label{DistI}
\end{equation}
where $c=(\sqrt{5}-1)/2 \approx 0.618$ is the golden ratio conjugate (or silver ratio), were also considered.  As before, we will denominate EAU, EAP, EAA and EAI to, respectively, each one of these versions of the Edwards-Anderson model.      

In order to simulate these models, we use a parallel tempering algorithm \cite{Geyer1991,Hukushima1996}. It consists in making an ensemble of $m$ replicas of the system, each of which is at temperature $T_k$ ($T_1 \geq T_k \geq T_m$). The basic idea of this algorithm is to simulate independently each replica with a standard Monte Carlo dynamics, and to swap periodically the configurations of two randomly chosen replicas. The purpose of this swap is to try to avoid that replicas at low temperatures get stuck in local minima. Thus the highest temperature, $T_1$, is set in the high-temperature phase where relaxation time is expected to be very short and there exists only one minimum in the free energy landscape. The lowest temperature, $T_m$, is set in the low-temperature phase. 

In order to implement this algorithm, we choose equally spaced temperatures, i.e. $T_k-T_{k+1}= \left(T_{1} - T_m \right)/(m-1)$.  Each replica is independently simulated by a single spin-flip dynamics where updates are attempted with a probability given by the Metropolis rule \cite{Metropolis}. On the other hand, the trial exchange of two configurations $X_k$ and $X_{k'}$ (corresponding to the $k$-th and $k'$-th replicas) is attempted and accepted with probability \cite{Hukushima1996}
\begin{equation}
W\left(X_k,\beta_k| X_{k'},\beta_{k'}\right)=\left\{
\begin{array}{cc}
1 & {\rm for}\ \ {\Delta \leq 0} \\
\exp(-\Delta)  & {\rm for}\ \ {\Delta>0},
\end{array}
\right. \label{exchange}
\end{equation}
where $\Delta=\left(\beta_{k'} - \beta_k \right)\left[ H(X_k) -H(X_{k'}) \right]$ and $\beta_k = 1/T_k$ (without loss of generality, we take the Boltzmann's constant equal to one).  A unit of time or parallel tempering step (PTS), consists of a number of $m \times N$ elementary spin-flip attempts followed by only one swap attempt.  As in Ref.~\cite{Hukushima1996}, we restrict the replica exchange to the case $k'= k+1$. 

The parallel tempering algorithm can be used as a heuristic to obtain GS configurations \cite{Moreno2003,Roma2009}.  For this application it is not necessary to reach equilibrium, because only low-energy configurations are sought.  Then, we have used this algorithm to calculate the GS energy.  As in reference \cite{Roma2009}, where only EAB and EAG models were studied, we have chosen $m=20$ and the extreme temperatures as $T_1=1.6$ and $T_m=0.1$.  In addition, the number of PTSs used here for the models with discrete (continuous) bond distributions, are the same ones that were used in reference \cite{Roma2009} to calculate the GS energy of the EAB (EAG) model.  

On the other hand, to calculate the GS entropy we have proceeded differently.  After an appropriate number of PTS, the parallel tempering algorithm allows to reach equilibrium and to calculate the mean energy at all temperatures.  As we will see in the following section, integrating this curve one can obtain a reliable value of the GS entropy.         

\section{\label{TIM}Thermodynamic integration method}

In the following, we briefly describe the thermodynamic integration method (TIM) \cite{Kirkpatrick1977,Binder1985,Roma2004}.  Given a model with a fixed number of entities (spins) $N$, we can write the basic relationship   
\begin{equation}
\frac{1}{T}=\left( \frac{\partial S}{\partial U} \right)_N,
\end{equation}
where $U(N,T)$ and $S(N,T)$ are the total internal energy and entropy, respectively.  Integrating this equation, the GS entropy of a sample $x$ is 
\begin{equation}
s_x(N,0)=s_x(N,T_R)+\int_{u_x(N,T_R)}^{u_x(N,0)} \frac{du_x}{T}. \label{Int0}
\end{equation}
Here, lower case letters denote quantities per spin and $T_R$ is the temperature of a reference state for which entropy is known.  While the integral in the second term can be accurately estimated by a Monte Carlo simulation, appropriate reference states are difficult to find. However, for an Ising spin system when $T_R \to \infty$, a trivial reference state with entropy      
\begin{equation} 
\lim_{T_R \to \infty} s_x(N,T_R) = \frac{\ln 2^N}{N}= \ln 2 , \label{RefEnt1}
\end{equation} 
is valid for any sample.  In practice, good results are obtained with equation (\ref{RefEnt1}) if the energy is calculated for a great number of high temperatures \cite{Roma2004}.  

Nevertheless, the TIM performance can be improved if a suitable reference entropy is determined at a not very high temperature.  Such calculation can be made by the high-temperature expansion method \cite{Bovier,Thouless1977,Kirkpatrick1978}. Let us consider the standard identity for Ising spin systems 
\begin{equation}      
\exp \left( \beta  J_{ij} \sigma_i \sigma_j \right) = \cosh \left(\beta  J_{ij} \right) \left[ 1+ \sigma_i \sigma_j \tanh \left(\beta  J_{ij} \right) \right]. 
\end{equation}      
Then, the partition function for a particular sample can be write as 
\begin{eqnarray} 
Z_x &=& \sum_\Omega \prod_{(ij)} \exp \left(\beta  J_{ij} \sigma_i \sigma_j \right) \nonumber \\
    &=& \sum_\Omega \prod_{(ij)} \cosh \left(\beta  J_{ij} \right) \left[ 1+ \sigma_i \sigma_j \tanh \left(\beta  J_{ij} \right) \right] \nonumber \\
    &=& 2^N \prod_{(ij)} \cosh \left(\beta  J_{ij} \right) \frac{1}{2^N} \sum_\Omega \prod_{(ij)} \left[ 1+ \sigma_i \sigma_j \tanh \left(\beta  J_{ij} \right) \right], 
\end{eqnarray}      
where the sum run over the $2^N$ configurations of the system.  Then, the free energy per spin, $f$, is
\begin{eqnarray}   
-\beta f &=& \frac{1}{N} \left[\ln Z_x \right]_{\mathrm{av}} \nonumber \\
         &=& \ln 2 + \frac{1}{N} \left[ \ln \left( \prod_{(ij)} \cosh \left(\beta  J_{ij} \right) \right)  \right]_{\mathrm{av}} + \nonumber \\ 
&& +\frac{1}{N} \left[ \ln \left( \frac{1}{2^N} \sum_\Omega \prod_{(ij)} \left[ 1+ \sigma_i \sigma_j \tanh \left(\beta  J_{ij} \right) \right] \right)  \right]_{\mathrm{av}}, \label{FreeEnergy}
\end{eqnarray}   
where $\left [ \cdots \right]_{\mathrm{av}}$ represent a disorder average. Given a probability bond distribution $P(J_{ij})$ and considering that hypercubic samples have $dN$ bonds, the second term in the right-hand side of equation (\ref{FreeEnergy}) can be write as
\begin{eqnarray}   
I &\equiv& \frac{1}{N} \left[ \ln \left( \prod_{(ij)} \cosh \left(\beta  J_{ij} \right) \right)  \right]_{\mathrm{av}}  \nonumber \\
&=& \frac{1}{N} \sum_{(ij)} \int_{-\infty}^{\infty} \ln \left[ \cosh \left(\beta  J_{ij} \right) \right] P(J_{ij}) dJ_{ij}  \nonumber \\
&=& d \int_{-\infty}^{\infty} \ln \left[ \cosh \left(\beta  J_{ij} \right) \right] P(J_{ij}) dJ_{ij} . \label{Int1}
\end{eqnarray}   
Neglecting the third term in the right-hand side of equation (\ref{FreeEnergy}), the free energy can be approximated by    
\begin{equation}
-\beta f \approx \ln 2 + I.
\end{equation}
From this equation, we can calculate the internal energy 
\begin{equation}
u = \frac{\partial \left(\beta f \right)}{\partial \beta} \approx  - \frac{\partial I}{\partial \beta}  \label{Energy}
\end{equation}
and entropy 
\begin{equation}
s = -\beta f + \beta u \approx \ln 2 + I - \beta \frac{\partial I}{\partial \beta}.  \label{Entropy}
\end{equation}
Then, the integral (\ref{Int1}) should be calculated for each model to determine the entropy of a suitable reference state. 

For the EAB model the integral (\ref{Int1}) is 
\begin{equation}
I_{\mathrm{B}} = d \ln \left[ \cosh \beta  \right], \label{IB1}
\end{equation}
and 
\begin{eqnarray}
u_{\mathrm{B}} \approx - d \tanh \beta   \label{uB1} \\
s_{\mathrm{B}} \approx \ln 2 + d \ln \left[ \cosh \beta  \right] -d \beta \tanh \beta  \label{sB1}  .  
\end{eqnarray}
It is easy to generalize this result for the case of the discrete bond distributions such as the $2p$-delta (\ref{DistP}),   
\begin{eqnarray}
u_{\mathrm{P}} \approx - \frac{d}{p} \sum_{i=1}^p ib \tanh \left( ib \beta \right)  \label{uP1} \\
s_{\mathrm{P}} \approx \ln 2 + \frac{d}{p} \sum_{i=1}^p \ln \left[ \cosh \left( ib \beta \right)  \right] -\frac{d \beta}{p} \sum_{i=1}^p  ib \tanh \left( ib \beta \right)  \label{sP1}  ,  
\end{eqnarray}
the asymmetric (\ref{DistA}), 
\begin{eqnarray}
u_{\mathrm{A}} \approx - \frac{d}{2} \left[ \tanh \left(\beta \right) + a \tanh \left(a \beta \right) \right]  \label{uA1} \\
s_{\mathrm{A}} \approx \ln 2 + \frac{d}{2} \ln \left[ \cosh \left(\beta \right) \cosh \left(a \beta \right)  \right] - \frac{d \beta}{2} \left[ \tanh \left(\beta \right) + a \tanh \left(a \beta \right) \right]  \label{sA1}  ,  
\end{eqnarray}
and the irrational (\ref{DistI}), 
\begin{eqnarray}
u_{\mathrm{I}} \approx - \frac{d}{2} \left[ \tanh \left(\beta \right) + c \tanh \left(c \beta \right) \right]  \label{uI1} \\
s_{\mathrm{I}} \approx \ln 2 + \frac{d}{2} \ln \left[ \cosh \left(\beta \right) \cosh \left(c \beta \right)  \right] - \frac{d \beta}{2} \left[ \tanh \left(\beta \right) + c \tanh \left(c \beta \right) \right]  \label{sI1}  .  
\end{eqnarray}
Notice that the two last models have the same reference state.  

For continuous bond distributions we have used the following Taylor expansion 
\begin{equation}
\ln \left[ \cosh \left(y \right) \right] = \frac{1}{2} y^2 - \frac{1}{12} y^4 + \frac{1}{45} y^6 - \frac{17}{2520} y^8 + \cdots.  \label{TaylorExp}
\end{equation}
To obtain a reference state for the EAG model, let us consider the integral 
\begin{equation}
F_n(\alpha)= \frac{1}{\sqrt{2 \pi}}  \int_{-\infty}^\infty \exp(- \alpha y^2) y^n dy. \label{IntF}
\end{equation}
It is easy to show \cite{Reif} that, for integer $n \ge 0$, $F_0(\alpha)=1/\sqrt{2 \alpha}$, $F_1(\alpha)=0$ and the following recurrence relation holds
\begin{equation}
F_n(\alpha)=- \frac{d}{d \alpha} \left[ \frac{1}{\sqrt{2 \pi}}  \int_{-\infty}^\infty \exp(- \alpha y^2) y^{n-2} dy \right] = - \frac{d F_{n-2}(\alpha)}{d \alpha}.
\end{equation}
Using the Taylor expansion (\ref{TaylorExp}) and the previous relation, we can approximate the integral (\ref{Int1}) by the first four terms      
\begin{equation}
I_{\mathrm{G}}=d \left[\frac{1}{2}  \beta^2 - \frac{1}{4} \beta^4 + \frac{1}{3} \beta^6 - \frac{17}{24} \beta^8 + \cdots \right],
\end{equation}
and the energy and the entropy can be expressed as 
\begin{equation}
u_{\mathrm{G}} \approx - d \left[\beta - \beta^3 + 2 \beta^5 - \frac{17}{3} \beta^7 \right] \label{uG1}
\end{equation}
and 
\begin{equation}
s_{\mathrm{G}} \approx \ln 2 - d \left[\frac{1}{2} \beta^2 - \frac{3}{4} \beta^4 + \frac{5}{3} \beta^6 - \frac{119}{24} \beta^8 \right], \label{sG1}
\end{equation}
respectively.  On the other hand, considering the Taylor expansion (\ref{TaylorExp}) and by a direct integration, for the continuous uniform distribution we have that 
\begin{equation}
I_{\mathrm{U}}=d \left[\frac{1}{2}  \beta^2 - \frac{3}{20} \beta^4 + \frac{3}{35} \beta^6 - \frac{17}{280} \beta^8 + \cdots \right],
\end{equation}
\begin{equation}
u_{\mathrm{U}} \approx - d \left[\beta - \frac{3}{5} \beta^3 + \frac{18}{25} \beta^5 - \frac{17}{35} \beta^7 \right]
\end{equation}
and
\begin{equation}
s_{\mathrm{U}} \approx \ln 2 - d \left[\frac{1}{2} \beta^2 - \frac{9}{20} \beta^4 + \frac{3}{7} \beta^6 - \frac{17}{40} \beta^8 \right].
\end{equation}

\begin{figure}[t]
\begin{center}
\includegraphics[width=\linewidth,clip=true]{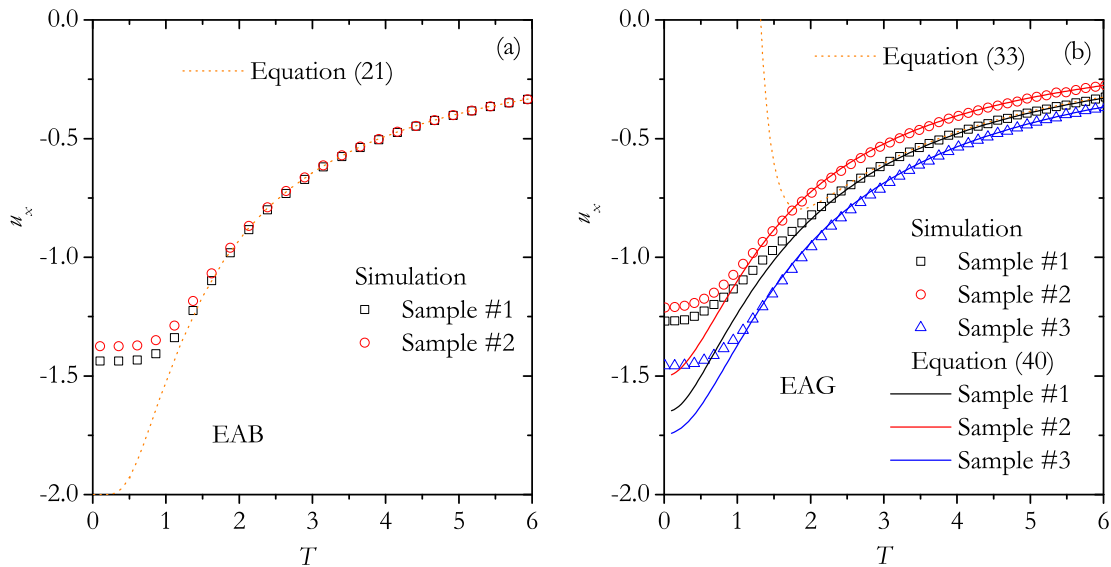}
\caption{The mean energy as function of $T$ for different 2D samples of $L=8$ and the corresponding analytical approximations (see text).  (a) The EAB model and (b) the EAG model. } \label{figure1}
\end{center}
\end{figure}

Previous reference states should be used carefully to calculate the GS entropy.  The main problem arises when the samples are generated \cite{Pazmandi}.  Let us consider for example the EAB model.  A canonical approach is implemented when samples are built with half of the bonds of each sign.  We refer to these as {\em canonical samples}.  On the other hand, if a grand canonical approach is used, {\em grand canonical samples} are generated in which bonds are put on each edge of the lattice with a probability given by the corresponding distribution (in this case $\pm 1$ bonds are chosen with equal probability).  Figure~\ref{figure1}(a), shows for the 2D EAB model, a comparison between the mean energy curves calculated with parallel tempering, for a canonical ($\#1$) and a grand canonical ($\#2$) samples (simulation parameters are given below). Here
\begin{equation}
u_x =\frac{ \langle H \rangle_T }{N}  \label{energy1}
\end{equation}  
is the mean energy, where $\langle \cdots \rangle_T$ represents a thermal average.  Although the lattice size $L=8$ is small, above $T \approx 4$ both curves match very well with the equation (\ref{uB1}).  Then, whatever the canonical or the grand canonical approach be used, an accurate GS entropy will be obtained using the reference state given by equation (\ref{sB1}) at this temperature.  This is due to that all bonds have the same magnitude and the hyperbolic cosine is a even function.  Therefore the integral (\ref{Int1}) is not sensitive to fluctuations in the bond's values.

A different situation arises for the systems for which only grand canonical samples can be generated.  Figure~\ref{figure1}(b) shows the mean energy for three distinct samples of the EAG model of size $L=8$.  Notice that the equation (\ref{uG1}) matches very well with the curve corresponding to sample $\#1$, but not with those of samples $\#2$ and $\#3$.  Then, the entropy (\ref{sG1}) only can be used as a reference state for sample $\#1$.  The problem is that, for this model, it is not possible to calculate a general low-temperature reference, because the integral (\ref{Int1}) is very sensitive to particular realizations of disorder.  Although these discrepancies disappear for bigger $L$, such samples are difficult to equilibrate at low temperatures.  Then, to extrapolate to the thermodynamic limit, we need to calculate the entropy for small lattice sizes.

A simple solution consists in to calculate a particular reference state for each sample.  Instead of performing a disorder average, we consider the complete free-energy expression 
\begin{equation}   
-\beta f_x = \frac{1}{N} \ln Z_x  \approx  \ln 2 + \frac{1}{N} \sum_{(ij)} \ln \left[ \cosh \left(\beta  J_c{ij} \right) \right],   \label{FreeEnergy2}
\end{equation}   
where again we have neglected the third term in the right-hand side of this equation.  Then, the mean energy and entropy equation (\ref{Energy}) and (\ref{Entropy}), respectively, are
\begin{equation}   
u_x \approx - \frac{1}{N} \sum_{(ij)}  J_{ij} \tanh \left(\beta  J_{ij} \right)    \label{Energy2}
\end{equation}   
and
\begin{equation}   
s_x \approx  \ln 2 + \frac{1}{N} \sum_{(ij)} \ln \left[ \cosh \left(\beta  J_{ij} \right) \right] - \frac{\beta}{N} \sum_{(ij)}  J_{ij} \tanh \left(\beta  J_{ij} \right) .   \label{Entropy2}
\end{equation}   
Given a specific sample, the sums in equations (\ref{Energy2}) and (\ref{Entropy2}) can be evaluated directly.  Figure~\ref{figure1}(b) shows the function (\ref{Energy2}) for the three samples of the 2D EAG model.  From this comparison, it is evident that equation (\ref{Entropy2}) will be a good entropy reference state for these samples. In fact, equations (\ref{Energy2}) and (\ref{Entropy2}) can be used for all, canonical and grand canonical samples.    

Although both approaches can be used for discrete bond distribution, in this work we have only studied canonical samples of the EAB and EAA models.  On the other hand, grand canonical samples were considered for the EAP, EAI and the models with continuous bond distribution, EAG and EAU.  In each case, to determine an appropriate reference temperature $T_R$, first we have calculated the exact GS entropy for several samples up to $L=8$, using a branch-and-bound algorithm \cite{Hartwig1984}.  Then, for these same samples the TIM, improved with a suitable reference as we discussed before, was used to calculate numerically each one of the corresponding GS entropies.  $T_R$ was chosen as the minimum temperature at which, for each one of the samples with $3 \le L \le 8$, the numerical estimation of the GS entropy agrees, within the simulation error, with the exact value.  

We calculate the integral (\ref{Int0}) between this reference temperature and $T_G$, a very low temperature close to $T=0$ (the lowest temperature in the parallel tempering should be close to but not exactly equal to zero). Table \ref{table1} shows, for each one of the models, the most important parameters that we have used in the our simulations: the temperatures $T_R$ and $T_G$, and the maximum lattice size studied, $L_{\mathrm{max}}$.  In all cases we have used $m=300$ replicas, $10^6$ PTSs (for samples with $L=L_{\mathrm{max}}$) and we have chosen the extreme temperatures in the parallel tempering as $T_1=T_R$ and $T_m=T_G$.  For lattice sizes $L<L_{\mathrm{max}}$, an equal or smaller number of PTSs were used.  

\begin{table}[h]
\begin{center}
\begin{tabular}{|c|c|c|c|}
\hline
Model & $T_R$ & $T_G$   &  $L_{\mathrm{max}}$      \\
\hline
EAB   & $7$   & $0.1$   &  $20$                    \\
EAG   & $5$   & $0.01$  &  $20$                     \\
EAU   & $6$   & $0.01$  &  $14$                      \\
EAP   & $7$   & $0.1$   &  $14$                      \\
EAA   & $5$   & $0.05$  &  $14$                          \\
EAI   & $9$   & $0.05$  &  $14$                           \\
\hline
\end{tabular}
\caption{\label{table1} Reference temperatures used to calculate the GS entropy and  maximum size studied for each model.}
\end{center}
\end{table}

\section{\label{NumRes} Numerical results}

In this section we present the main results of our simulations.  For each model, the disorder average for the maximum lattice size was performed over $10^3$ samples while, for smaller sizes, up to $10^5$ samples were necessary to obtain accurate values of the GS energy and entropy.  

To extrapolate, we have fitted our data of the GS energy to this scaling function
\begin{equation}
f_u=u_\infty + g_u L^{-d_u}, \label{function_u}
\end{equation}
where $u_\infty$ is the GS energy value in the thermodynamic limit and $g_u$ and $d_u$ are two parameters.  A similar scaling function 
\begin{equation}
f_s=s_\infty + g_s L^{-d_s} , \label{function_s}
\end{equation}
with parameters $g_s$ and $d_s$, was used to estimate the thermodynamic limit of the GS entropy, $s_\infty$.  For a given fit, the corresponding goodness-of-fit parameter $Q$ is calculated \cite{numerical}. A value $Q \gtrsim 0.1$ is considered as indication of good quality of the fit. As is usually the case, scaling functions do not include all possible finite size corrections, and therefore better fits are obtained when data for very small sizes are left out. On the other hand, leaving out too many points can result in large error bars for the best fit parameters. Then, the results presented here were obtained by fitting the data over the largest range that gives a goodness-of-fit of $Q \gtrsim 0.1$. 

\subsection{EAB and EAG models}

\begin{figure}[t]
\begin{center}
\includegraphics[width=\linewidth,clip=true]{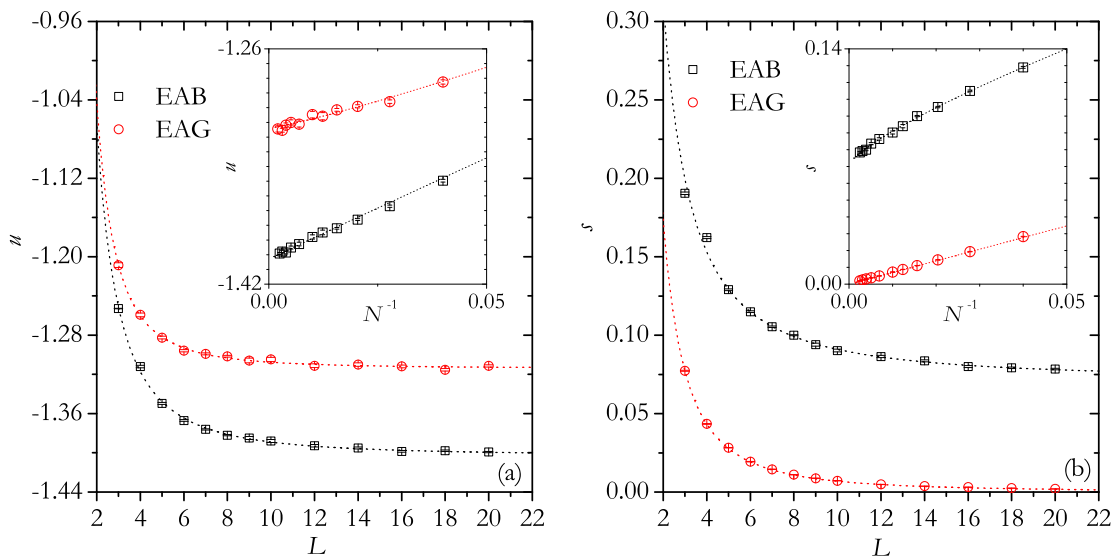}
\caption{The GS (a) energy and (b) entropy versus $L$, for the EAB and EAG models.  The dotted lines are the best fits obtained. The insets show, for both models, the GS energy and entropy as function of $N^{-1}$.} \label{figure2}
\end{center}
\end{figure}

Figure \ref{figure2} (a) shows the GS energy for the EAB and EAG models and different lattice sizes.  The results of fitting these curves with the scaling function (\ref{function_u}) are presented in table \ref{tableU}.  Although (for these models) the maximum lattice size simulated are not very large, the values of $u_\infty$ agree very well with the most accurate results reported in previous works: $u_\infty=-1.40193(2)$ \cite{Palmer1999}, $u_\infty=-1.40197(2)$ \cite{Campbell2004} and $u_\infty=-1.4009(3)$ \cite{Roma2009} for the EAB model; $u_\infty=-1.31479(2)$ \cite{Campbell2004} and $u_\infty=-1.3149(5)$ \cite{Roma2009} for the EAG model.  Only the exponent $d_u=1.95(11)$ for EAB model is a little different from $d_u = 2.13(4)$, the value that we have obtained previously \cite{Roma2009}.  The reason for this discrepancy is that the fit of the GS energy with equation (\ref{function_u}), is very sensitive to the range of $L$ used (in reference \cite{Roma2009} was used a range of $5-30$).

The GS entropy for these same models is shown in figure \ref{figure2} (b), while in table \ref{tableS} are presented the results of fitting these curves with the scaling function (\ref{function_s}). The value of $s_\infty=0.0714(9)$ for the EAB model is close to the results reported in the literature: $s_\infty=0.075(2)$ \cite{Saul1994}, $s_\infty=0.0709(6)$ \cite{Zhan2000} and $s_\infty=0.078(5)$ \cite{Hartmann2001b}. On the other hand, because the EAG model has a non-degenerated fundamental level (that has only two configurations related by a global spin-flip), the thermodynamic limit of the GS entropy is expected to be zero.  However, we obtain $s_\infty=0.0003(1)$, a value very close to, but not zero. This is due to the existence of systematic errors in the implementation of the thermodynamic integration method \cite{Roma2004}. Among others, one source of error is the (bad) assumption of that the temperature $T_G$ is sufficiently low, so as not to affect the calculation of the integral (\ref{Int0}). To improve our result, we should equilibrate each sample up to a lower temperature.  As this is very hard to do, we assume that the entropy values are affected by a systematic error of order $10^4$. This shows the accuracy of our implementation of the thermodynamic integration method, and justifies why a system with a trivial GS as the EAG model was studied.  Finally, in table \ref{tableS} we can see that the exponent $d_s$ for this model is very close to $2$.  This is correct because the exact GS entropy is simply 
\begin{equation}
s=\frac{1}{N}\ln 2 .  \label{GSentropy}
\end{equation}  
The inset in figure \ref{figure2} (b) shows that this equation and the GS entropy for the EAG model agree very well for all sizes.   

\subsection{EAU and EAP models}

\begin{figure}[t]
\begin{center}
\includegraphics[width=\linewidth,clip=true]{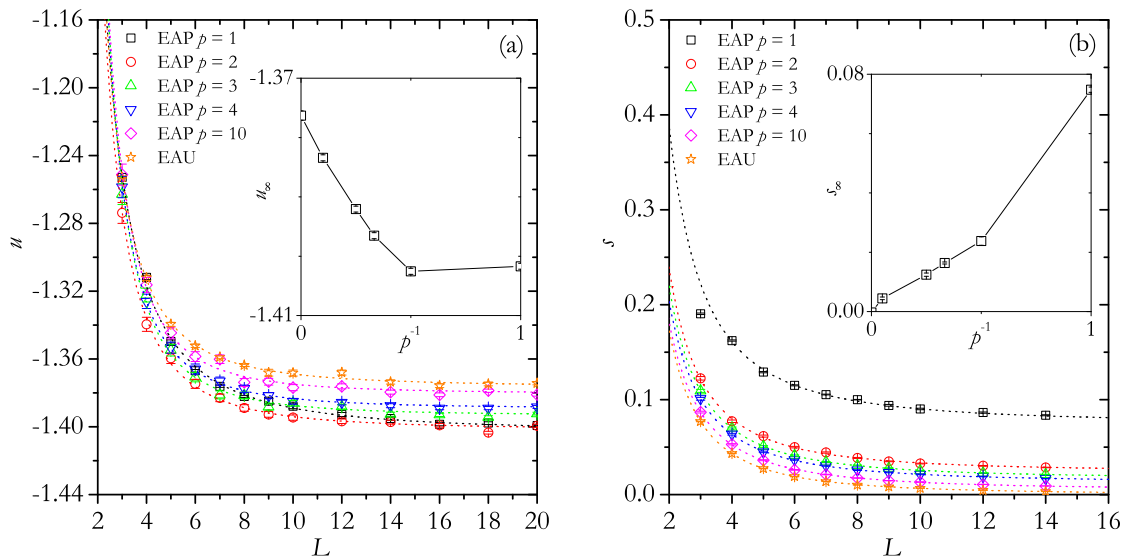}
\caption{The GS (a) energy and (b) entropy versus $L$, for the EAU model and the EAP model with different values of $p$.  The dotted lines are the best fits obtained. The insets show the GS energy  $u_\infty$ and entropy $s_\infty$ as function of $p^{-1}$.} \label{figure3}
\end{center}
\end{figure}

Figure \ref{figure3} (a) shows the GS energy for the EAU and EAP models for different lattice sizes and $p$ values.  The inset shows the dependence with the parameter $p$ of the GS energy $u_\infty$, which is obtained by fitting the data with the scaling function (\ref{function_u}).  We can see that for $p \ge 2$, the curve monotonically tends to the corresponding value for the EAU model (see table \ref{tableU}).  The same is observed in figure \ref{figure3} (b) for the GS entropy.    

By fitting the entropy curve of the EAU model, we obtain a negative value $s_\infty=-0.0003(1)$.  As this model has a continuous bond distribution (\ref{DistU}), it should have a non-degenerated fundamental level with entropy zero. Again, the problem is the systematic error affecting the results produced by the thermodynamic integration method. Nevertheless, figure \ref{figure3} (b) shows that the GS entropy calculated with this technique for each lattice size follows the exact curve (\ref{GSentropy}).

\subsection{EAA model}

\begin{figure}[t]
\begin{center}
\includegraphics[width=\linewidth,clip=true]{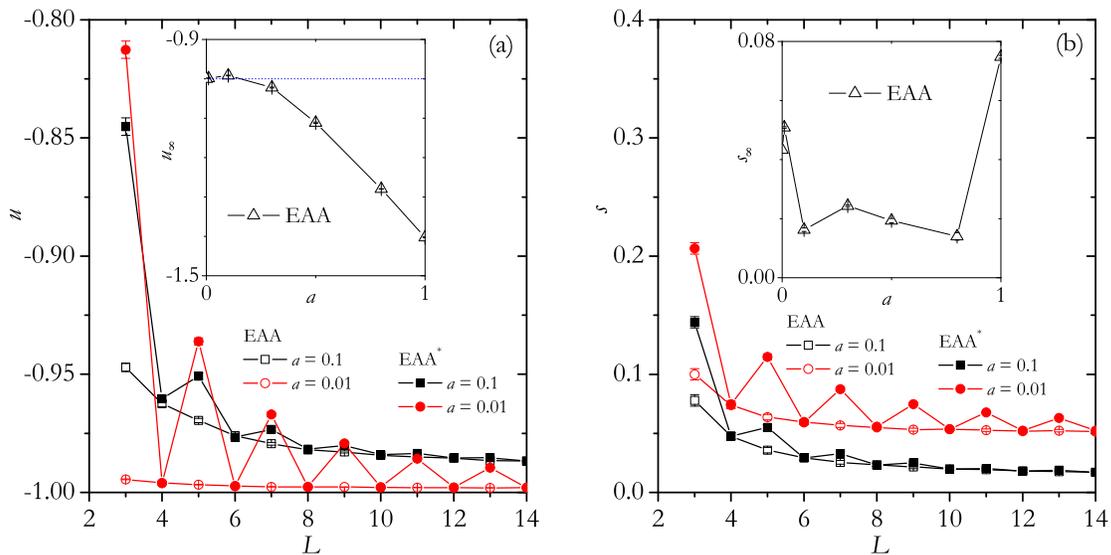}
\caption{The GS (a) energy and (b) entropy versus $L$, for the EAA and EAA$^*$ models, for two values of $a$ as indicated.  The dotted lines are the best fits obtained. The insets show the GS energy  $u_\infty$ and entropy $s_\infty$ as function of the parameter $a$.} \label{figure4}
\end{center}
\end{figure}

In the EAA model, bonds are chosen randomly with equal probability between two values: $+1$ (ferromagnetic bond) and $-a$ (antiferromagnetic bond).  We could have defined a similar distribution by exchanging the signs of these bonds,
\begin{equation}
P_{\mathrm{A}^*}(J_{ij})=\frac{1}{2} \left[\delta(J_{ij}-a)+\delta(J_{ij}+1)\right].   \label{DistA2}
\end{equation}
We will call this the EAA$^*$ model.  Figures \ref{figure4} (a) and (b) show, respectively, the GS energy and entropy for the EAA and EAA$^*$ models for two different values of parameter $a$.  The most outstanding behavior is that both quantities seem to oscillate for the EAA$^*$ model.  In particular, the GS energy and entropy for the two models, do not match for odd values of the lattice size.  This discrepancy increases with decreasing $a$, but tends to disappear with increasing $L$.     

The reason of this behavior is in the geometry of the lattice, which is square, and the periodic-periodic boundary conditions.  For small values of $a$, antiferromagnetic bonds in the EAA model are weak and the GS is formed by ferromagnetic islands. This structure is not affected considerably by the $L$ value: ferromagnetic clusters can fill the lattice with an even or odd $L$.  However, for the EAA$^*$ model the GS is antiferromagnetic and a different situation arises.  Now, due that nearest-neighbors spins point in opposite directions, the GS configurations are characterized by the homogeneous magnetization of both sublattices but with different orientations.  The GS energy and entropy are affected by the periodic-periodic boundary conditions, because to accommodate a large (percolating) antiferromagnetic island in a lattice with odd $L$, it is necessary to create an energy wall of the same length order. This does not happen for lattices with even $L$, for which the antiferromagnetic structure can fill the system. As the defect energy (or entropy) depends approximately on $L$ and the GS energy (or entropy) depends on $L^2$, the discrepancies disappear with increasing size and, in the thermodynamic limit, both models will have the same $u_\infty$ ($s_\infty$).

In order to carry out the fitting, we have only used the EAA data.  Insets in figures \ref{figure4} (a) and (b) show, respectively, the dependence with parameter $a$ of the GS energy $u_\infty$ and entropy $s_\infty$. For $a=1$, the EAA and EAB models are equivalent and therefore have the same values of energy and entropy.  When $a=0$ half of the bonds are zero and the other half $-1$, and then the GS energy is $\lim_{a \to 0} u_\infty=-1$.  The GS entropy is also easy to calculate: since the probability that one spin has four bonds of zero strength is $(1/2)^4=1/16$, in the thermodynamic limit the number of {\em free} spins (those spins whose flipping does not change the energy of the state) will be $n=N/16$. Then 
\begin{equation}     
\lim_{a \to 0} s_\infty=\frac{1}{N} \ln\left(2^n \right)= \frac{\ln 2}{16} \approx 0.0433.
\end{equation}     

The insets in figures \ref{figure4} (a) and (b) show that the GS energy changes smoothly between these limits but the GS entropy does not.  The behavior of $s_\infty$ can be explained considering that the number or free spins depends strongly on the parameter $a$. For $a=1$ all bonds have the same magnitude (i.e. $\pm 1$) and spins with two frustrated and two satisfied bonds will be free. If the parameter diminishes a little, $a \lesssim 1$, $n$ changes a lot: in order to have a free spin now it is necessary that, either four bonds converging onto it have the same magnitude (with two of them being frustrated and two satisfied), or that two of them have magnitude $1$ and the other two $a$ (with two bonds of magnitude $1$ and $a$ frustrated and the other two bonds satisfied).  Because the probability of this happening is smaller than before, the GS entropy falls abruptly [see inset in figure \ref{figure4} (b)].  A new possibility arises when $a=1/3$: in a GS configuration, a spin with three satisfied (frustrated) bonds of magnitude $a$  and one frustrated (satisfied) bond of magnitude $1$, will be free.  Then, the GS entropy should increase a little at $a=1/3$.           

Another characteristic of the EAA model, is how the energy gap $\Delta H_0$ between the GS and the lowest excitation state depends on the parameter $a$.  Supposing that these excitations are due to single-spin flips only, at $a=1$ the gap is $\Delta H_0=4$ and this corresponds to flipping a spin with three bonds satisfied and one frustrated. Also, for $1/3 \le a <1$ is $\Delta H_0=2(1-a)$ [this gap correspond, for example, to spins with three bonds of magnitude $a$, with two of them being frustrated and one satisfied, and the remaining bond of magnitude $1$ satisfied], for $1/5 \le a < 1/3$ is $\Delta H_0=2(1-3a)$ [spins with three frustrated bonds of magnitude $a$ and one satisfied bond of magnitude $1$] and for $a < 1/5$ is $\Delta H_0=4a$ [spins with four bonds of magnitude $a$, with three of them being satisfied and one frustrated]. Interestingly, if we consider excitations in which are involved many spins, for $a<1$ there are a few samples with a smaller gap.  For example, at $a=0.1$ the gap should be $\Delta H_0=0.4$, but we have found samples of size $L=10$ with $\Delta H_0=0.2$: excitations are due to {\em droplets} whose walls are formed by a {\em net number} of $11$ satisfied bonds of magnitude $a$ and one frustrated bond of magnitude $1$.  Nevertheless, as the parameter $a$ is a rational number, the energy levels of this model form a discrete spectrum.

\subsection{EAI model}

\begin{figure}[t]
\begin{center}
\includegraphics[width=\linewidth,clip=true]{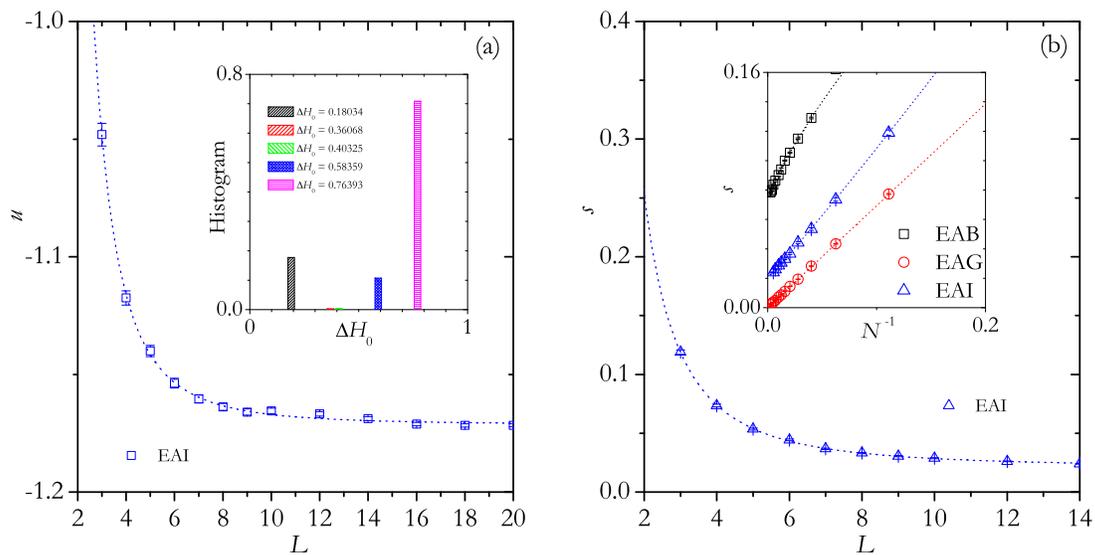}
\caption{The GS (a) energy and (b) entropy versus $L$ for the EAI model. The dotted lines are the best fits obtained. The inset in (a) shows the histogram of the energy gap obtained for $10^3$ samples of $L=10$.  The inset in (b) shows a comparison between the GS entropy $s_\infty$ of the EAI, EAB and EAG models, as function of $N^{-1}$.} \label{figure5}
\end{center}
\end{figure}

Figures \ref{figure5} (a) and (b) show the GS energy and entropy for the EAI model and different values of $L$.   The results of fitting these curves with the scaling functions (\ref{function_u}) and (\ref{function_s}) are presented, respectively, in tables \ref{tableU} and \ref{tableS}.  

Due that the parameter $c$ in the bond distribution (\ref{DistI}) is an irrational number, it is expected that the energy levels form a dense spectrum \cite{Amoruso2003}. We have found evidence that indicates that this is correct. Inset in figure \ref{figure5} (a) shows a histogram of the energy gap for the EAI model obtained for $10^3$ samples of $L=10$.  Although most of the samples have a $\Delta H_0 \approx 0.76393$ (corresponding to excitations of a single-spin flip), a considerable number of them have a smaller energy gap (corresponding to big droplets).     

Finally, we can see that contrary to the EAG and EAU models, which have also a dense spectrum of energy levels, here the GS entropy is not zero.  Inset in figure \ref{figure5} (b) shows a comparison between the curves of $s_\infty$ versus $N^{-1}$, for the EAI, EAB and EAG models.  It is evident that the GS of the EAI model is degenerated. An exponential number of GS configurations exists because the bond distribution $P_{\mathrm{I}}$ is discrete.  As the magnitude of bonds are similar to those of the $P_{\mathrm{A}}$ distribution with $a = 0.618$, the number of free spins and the GS entropy for both models should be similar.  We obtain $s_\infty = 0.0209(5)$ for the EAI model, which are very close to $s_\infty \approx 0.017$ for the EAA model (with $a = 0.618$).          

\begin{table}[t]
\begin{center}
\begin{tabular}{|c|c|c|c|c|}
\hline
Model  &  $u_\infty$     &   $g_u$   &   $d_u$    &  Range of $L$    \\
\hline
EAB    &  $-1.4024(10)$  & $1.2(2)$  & $1.95(11)$ &  $5-20$ \\
EAG    &  $-1.3136(13)$  & $1.5(4)$  & $2.39(22)$ &  $4-20$ \\
EAU    &  $-1.3763(8)$   & $1.6(1)$  & $2.32(6)$  &  $3-20$ \\
EAI    &  $-1.1713(6)$   & $1.6(2)$  & $2.78(7)$  &  $3-20$ \\
\hline
\end{tabular}
\caption{\label{tableU} Best fit parameters for the scaling function (\ref{function_u}) and the range of $L$ used. }
\end{center}
\end{table}

\begin{table}[t]
\begin{center}
\begin{tabular}{|c|c|c|c|c|}
\hline
Model  &  $s_\infty$     &   $g_s$    &   $d_s$    &  Range of $L$    \\
\hline
EAB    &  $0.0714(9)$    & $0.71(7)$  & $1.56(6)$  &  $5-20$ \\
EAG    &  $0.0003(1)$    & $0.694(7)$ & $2.00(1)$  &  $3-20$ \\
EAU    &  $-0.0003(1)$   & $0.70(8)$  & $2.01(1)$  &  $3-14$ \\
EAI    &  $0.0209(5)$    & $1.01(4)$  & $2.12(3)$  &  $3-14$ \\
\hline
\end{tabular}
\caption{\label{tableS} Best fit parameters for the scaling function (\ref{function_s}) and the range of $L$ used. }
\end{center}
\end{table}

\section{\label{Conc} Conclusions}

In this work we have used a parallel tempering algorithm to determine both the GS energy and, by means of the thermodynamic integration method, the GS entropy of the 2D Edwards-Anderson spin-glass model with different bond distributions.   

To implement the thermodynamic integration technique, we have built reference states by the method of high-temperature expansion. Although different strategies can be used for canonical and grand canonical samples a simple solution, consisting in to calculate a particular reference state with equations (\ref{Energy2}) and (\ref{Entropy2}) for each sample, works in all the cases.  By using this method, we have been able to calculate accurate values of the GS entropy.  This allowed us to make an study of six versions of the 2D Edwards-Anderson spin-glass model, which have different GS properties.

\section*{acknowledgments}
A.J. Ramirez-Pastor and F. Rom\'a acknowledge support from CONICET (Argentina) under Projects No. PIP112-200801-01332 and No. PIP114-201001-00172, and the National Agency of Scientific and Technological Promotion (Argentina) under Projects No. 33328 PICT-2005 and No. 2185 PICT-2007.


\end{document}